# Charge gradients around dendritic voids cause nanoscale inhomogeneities in liquid water

Tereza Schönfeldová[1], Nathan Dupertuis[1], Yixing Chen[1], Narjes Ansari[2], Emiliano Poli[3], David M. Wilkins[4], Ali Hassanali[3], Sylvie Roke[1,#]

[1]Laboratory for fundamental BioPhotonics (LBP), Institute of Bio-engineering (IBI), and Institute of Materials Science (IMX), School of Engineering (STI), and Lausanne Centre for Ultrafast Science (LACUS), École Polytechnique Fédérale de Lausanne (EPFL), CH-1015, Lausanne, Switzerland,

[2]Italian Institute of Technology, Via Morego 30, 16163 Genova, Italy

[3]The Abdus Salam International Center For Theoretical Physics, Condensed Matter and Statistical Physics (CMSP)

[4]Atomistic Simulation Centre, School of Mathematics and Physics, Queen's University Belfast, Belfast BT7 1NN, Northern Ireland, United Kingdom

[#]ahassana@ictp.it ; sylvie.roke@epfl.ch

**ABSTRACT**

Water is the matrix of life and is generally considered a homogeneous, uniform, liquid. Recent experiments report that water is instead a two-state liquid. However, subsequently, these findings were contested. The structure of water and whether we should think of it as uniform therefore remains an open question. Here, we report femtosecond elastic second harmonic scattering (fs-ESHS) of liquid water in comparison to an isotropic liquid ($CCl_4$) and show that water is indeed a non-uniform liquid, The coherent fs-ESHS intensity was interpreted, using molecular dynamics simulations, as arising from charge density fluctuations and consequentially enhanced nanoscale polarizabilities around transient voids having an average lifetime of 300 fs. Although voids were also present in $CCl_4$, they were not characterized by hydrogen bond defects and did not show strong polarizability fluctuations, leading to fs-ESHS of an isotropic liquid. The voids increased in number at higher temperatures above room temperature, in agreement with the fs-ESHS results. The measured nanoscale-femtosecond inhomogeneities therefore do not necessarily relate to the proposed two state model of water, but instead underscore the elusive nature of liquid water, and undoubtedly have an impact on any type of transition that occurs in liquid water.

**MAIN TEXT**

**Introduction**

Water is an enigmatic liquid. The structure of liquid water is most commonly represented as isotropic and uniform. Recent X-ray spectroscopic scattering measurements have suggested that this may be an over-simplification [1–6]. Based on fluctuations in the measured structure factor in combination with an analysis based on molecular dynamics (MD) simulations [7–15] water is proposed to be a 2-state liquid, composed of high-density locally disordered and low-density locally tetrahedrally ordered patches. The extent of the high and low density regions undergoes its most significant changes upon supercooling [16,17]. A recent theoretical study by the Goddard group refers to liquid water as a 'dynamic polydisperse polymer' consisting of water molecules with only two intact hydrogen bonds [18]. Molecular dynamics simulation studied by others, however, severely contest these interpretations and insist that water is a homogeneous liquid [7,19,20]. To understand the structure of water on the molecular and nanoscale level, there is a need for alternative experimental techniques that can probe water structure over nanometric length scales and, correspondingly, on time scales comparable to the femtosecond restructuring time of the hydrogen bond network of water. However, although there are many spectroscopic techniques to probe water structure [21–26] nearly all methods are sensitive to the hydration shell (sub-nanometric length scales involving at most a few layers of water) and spatiotemporal structural averaging, severely limiting our understanding of the nanometric length and femtosecond timescales on which water structures and transforms [27].

Recently, high-throughput femtosecond elastic second harmonic scattering (fs-ESHS, experimental details given in the SI, S1) has been invented [28,29]. Illustrated in Figure 1A, the interaction of near-infrared femtosecond laser pulse with a liquid can produce second harmonic (SH) photons that report on specific nanoscale (~> 1 nm) structuring of a liquid. Simple aqueous electrolyte solutions were shown to both incoherently and coherently emit SH photons [30]. The incoherent emission is known since the 1960s as Hyper Rayleigh scattering (HRS) and arises from an isotropic distribution of anisotropic molecules [31–33], which has been interpreted as the average structure of a liquid. The coherent emission was previously not associated with liquid structure [31–33]. It reports on a non-isotropic distribution of anisotropic molecules, and is sensitive to relatively long-range nanometer size effects thanks to the emitted electromagnetic field's cubic distance dependence between the correlated scatterers [30]. These correlations represent dynamic fluctuations over nanometer length scales in the polarizability of the liquid. In the case of dilute electrolyte solutions the length scale of fluctuations varied from ~6 to ~22 nm, corresponding to ~19 and ~70 hydration shells respectively. Coherent and incoherent emission are distinguishable by using different polarization combinations of the in- and out-going light [30]. In contrast to linear light, electron, x-ray or neutron scattering, fs-ESHS is background free, and when the coherent response is created, only those water molecules that are not averaged out in terms of their orientational distribution within a length scale on the order of ~1/10th of the emitted wavelength are reported on.

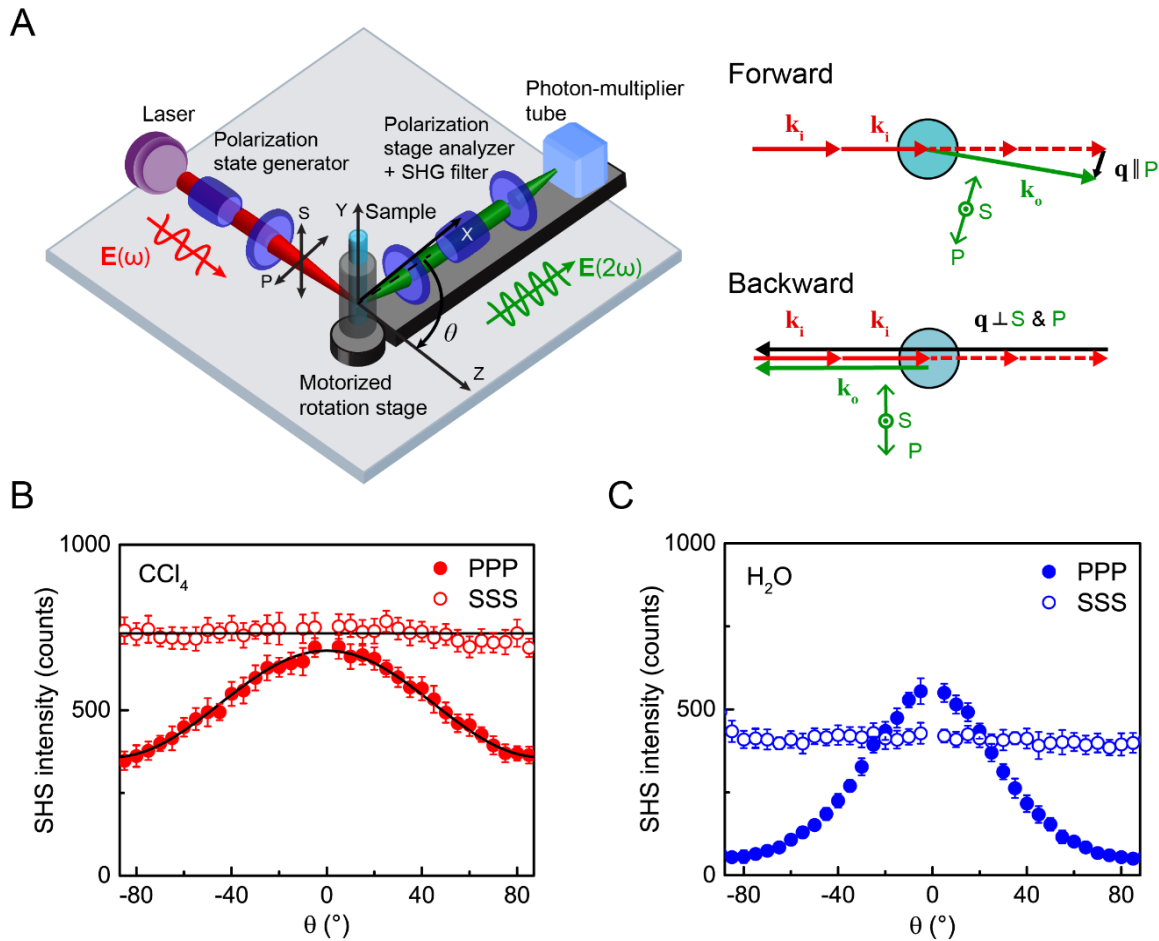

**Figure 1: Nanoscale femtosecond heterogeneities in water.** A: Illustration of the fs-ESHS experiment, together with drawings of the emitted (**k**) and scattering (**q**) wave vectors involved in forward and backward scattering. B, C: fs- ESHS patterns of neat $CCl_4$ (B) and neat $H_2O$ (C) collected at room temperature, using beams with all electromagnetic fields oscillating in the horizontal scattering plane (PPP) and the orthogonal vertical plane (SSS). The lines in panel B represent the fs-ESHS emission of isotropic liquid (S2).

Here, fs-ESHS measurements of liquid water in comparison to an isotropic liquid ($CCl_4$) are performed that show that water is a non-homogeneous liquid on femtosecond timescales. The SH emission of water is found to be partially coherent and arises from second-order polarization changes that originate from orientationally correlated water molecules. The SH emission of $CCl_4$, on the other hand, is completely incoherent, and originates from isotropically distributed molecules. MD simulations of water and a Lennard-Jones (LJ) liquid (similar to $CCl_4$) demonstrate the abundance of cavities or voids with an average life time of 300 fs. In both liquid water and the LJ liquid, space can be cataloged in terms of the presence of spherical and dendritic shaped voids [34,35]. Investigating electronic charge density distributions around the voids, we find that these are present for water, but not for the LJ liquid. In water, charge density fluctuations around the voids arise from the presence of coordination water defects in the hydrogen bond network [36]. This network is characterized by an enhanced nanoscale polarizability as computed by quantum chemistry calculations. We propose that these features are responsible for the coherent SH emission, and verify this further by temperature-dependent fs-ESHS experiments that show an increase in the void density at a higher temperature, in agreement with the MD predictions.

**Results and Discussion**

**fs-ESHS of water and $CCl_4$.** Figure 1 shows SH scattering patterns of neat $CCl_4$ (Fig. 1B) and neat $H_2O$ (Fig. 1C) that are collected at room temperature (295 K), using optical beams with all electromagnetic fields oscillating in either the horizontal scattering plane (PPP) and the orthogonal vertical plane (SSS). The first polarization combination reports on both coherent and incoherent SH scattering while the second polarization combination reports only on the incoherent SH scattering [30]. The patterns of $CCl_4$ display SH intensity in both SSS and PPP polarization combinations. Close to the forward scattering direction the intensity for both polarizations is equal. Note that it is not possible to measure at exactly 0° due to the presence of the strong fundamental beam. Such an equal intensity is a hallmark of an isotropic liquid and can well be modeled by HRS theory [32,33,37] as explained fully in section S2 of the SI. Fits to the incoherent HRS model are shown in Fig. 1B (black lines). The scattered SH light arises from electron density or polarizability fluctuations within single $CCl_4$ molecules that act as uncorrelated point sources. The same experiment is performed for liquid $H_2O$ (Fig. 1C). Interestingly, liquid water shows remarkably different behavior: The SH intensity in the forward direction is stronger for the PPP than the SSS polarization combination. Such behavior points to one of two things: Non-homogeneity in the water structure on a ~100 fs timescale, or an experimental artifact, as previously believed [38,39]. The latter can be excluded by the $CCl_4$ liquid data of Fig. 1B, which was recorded with the same experimental setup and the same experimental parameters. Additional evidence for the exclusion of the possibility of an artifact was obtained from two-photon fluorescence (2PF) and Hyper Raman measurements for both liquids (Fig. S1) as well as the difference between elastic and inelastic scattering (Fig. S2). This date is provided in section S3 of the SI.

The non-homogeneity in the structure and the presence of orientational correlations between the water molecules on a fs timescale follows directly from the measured intensities. Namely, the PPP polarization combination has a larger intensity than the SSS polarization combination in the near-forward scattering direction ($q \rightarrow 0°$). In this **q**-vector direction, the scattering wave vector **q** is positioned in the horizontal plane and is nearly perpendicular to the wave vector of the SH light, **k₀** (Fig. 1A), and parallel to the emitted SH electromagnetic field. When the emitted SH electromagnetic field oscillates parallel to **q**, it is possible to transfer momentum, but when the emitted SH electromagnetic field oscillates perpendicular to **q** it is not possible to transfer momentum [37]. Therefore the SSS intensity only probes the structure of individual water molecules while the PPP intensity can also be enhanced by their orientational correlations. This simple explanation can be verified by measuring the scattered SH light in the backward scattering direction (Fig. 1A). In this direction, **q** is parallel to the wave vector of the SH light, and perpendicular to both the S and P directions of the electromagnetic SH field. No difference is thus expected in the scattered light intensity for either S or P polarized SH light, as neither probes structural orientational correlations, even if they are present. Table S1 (SI, section S4) shows the results obtained.

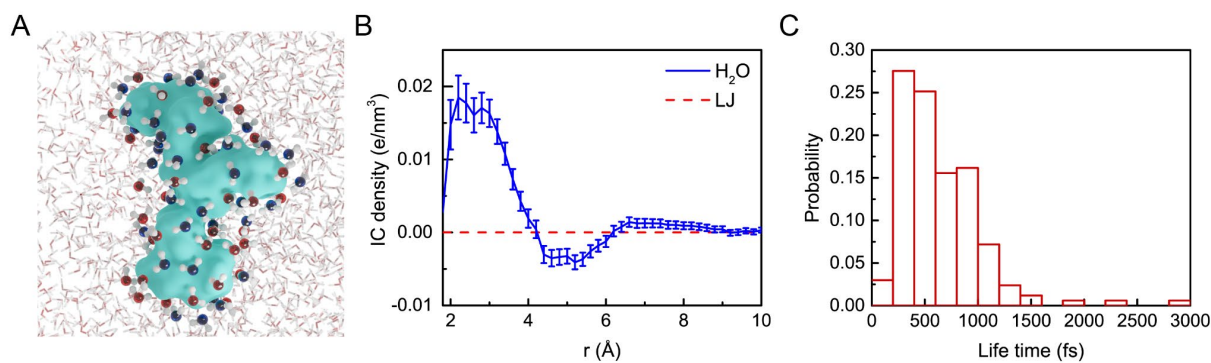

**Figure 2: Charge density fluctuations around transient nanoscale voids.** A: Illustration of a dendritic shaped void found in water. The color of the oxygen atom of the water molecule represents the total charge of the water molecule, where red shows positive and blue shows negative charge. B: The integrated charge density of water molecules as a function of distance from the surface surrounding the void. Interestingly, we observe a positive layer of charge around the void followed by a negative region before observing charge neutrality. The LJ liquid does not show such fluctuations. C: Distribution of the lifetime of voids seen in the simulations. The lifetime is examined starting from various initial conditions, and quantified by the time it takes for the void to disappear, or in other words, for a water molecule to enter the void.

**Nanoscale dendritic voids as possible source?** To observe such a pronounced coherent fs-ESHS response there has to be an abundant structural non-uniformity leading to orientational correlations in liquid water at room temperature with a lifetime > 100 fs. Orientational correlations are most pronounced around closed surfaces, and this effect has been used frequently as a way of interrogating the molecular structure of liquid water around nano-objects [28,40,41]. Furthermore, the polarization combination (PPP > SSS) is consistent with orientational correlations around a closed surface, and inconsistent with domains that have a specific directionality. The SH intensity depends strongly on size (having an intensity dependence of (size)$^6$ for small objects [42]), and shape (with more asymmetric shapes leading to bigger responses [43]), and the response is therefore expected to derive from structures that are larger than a few hydration shells (> 1 nm). One factor that possibly gives rise to these features is the transient variation of empty space in the liquid, which constitutes closed surfaces on a nanometer scale. These voids are also highly asymmetric in shape. Figure 2A shows an example of a void in water, extracted from the MD simulations, which were shown to be present in a wide class of different water models [35]. SI section S3-S4 provides the details of how the voids were computed and how their presence across a range of different water models was validated. There are two types of voids: small spherical voids that cover short length scales (< 0.5 nm), and dendritic shaped voids that are more delocalized, covering length scales of up to 2 nanometers. Although the voids are shown to exhibit local density fluctuations that might be relevant to explain the oscillations in the structure factor of SAXS [3] and X-Ray emission spectroscopy (XES) data [4], density fluctuations alone do not necessarily lead to the fs-ESHS emission, and both observables aren't necessarily linked.

A coherent fs-ESHS emission from a single object requires the following features: (1) A significant change in charge density or polarization and (2) A lifetime of the voids on the order of at least ~100 fs, the time scale of the probing laser pulses. To elaborate these possibilities, we employed linear-scaling density functional theory (LS-DFT) allowing for electronic structure calculations of large periodic systems consisting of 1000s of water molecules. From these simulations, density derived electrostatic charges (DDEC) are extracted and charge profiles around the voids examined (see SI section S7 for more details). The computed integrated charge density around voids is shown in Fig. 2B. For bulk $H_2O$ (blue trace), there is a significant change in charge density, which leads to charge gradients around the voids. A Lennard-Jones liquid (red dashed trace), representative of $CCl_4$, does have voids but does not display charge density fluctuations. The corresponding computed lifetime of the voids is on the order of several 100s of femtoseconds, with the mean being at 300 fs, as shown in Fig. 2C. Thus, the two structural requirements for observing a coherent fs-ESHS response from a dendritic void are met.

**Experimental and theoretical verification.** The charge density oscillations around the voids do not directly relate to the measured polarizability. To provide a more direct link between the fs-ESHS experiment and theory it is necessary to consider the electronic interferences of different molecules to explicitly account for correlations and compute nanoscale polarizabilities of transient structures. However, up until now there were only theories and computations that described either single molecules (quantified by the molecular hyperpolarizability) or macroscopic bulk media (which take the average of this quantity) [44]. Both approaches neglect the interferences, which are responsible for the measured orientational correlations. In order to compute the full SHS response form liquid water electronic interferences between all molecules will have to be computed on a length scale comparable to the wavelength. With the current state of the art of computational methods this is not achievable. Nevertheless, to account for electronic interferences on the length scale of voids, we utilize high level quantum chemistry calculations as detailed in section S8 of the SM, and apply them to clusters of water molecules containing up to 30 water molecules. Figure 3 shows several different voids composed of 18, 20, 25 and 30 water molecules (Fig. 3A), together with the computed SH intensity (Fig. 3B), which is the absolute square of the calculated nanoscale void polarizability. The intensity is normalized by the number of water molecules per void and plotted for the various sized voids so that it can be compared to the SH intensity of a single water molecule (SW). Strikingly, the intensity that arises per water molecule in the hydrogen bond network around the void is an order of magnitude larger than that of an isolated water molecule. Although other hydrogen bond network structures such as local coordination defects, rings or chains can also serve as potential scatterers, their contribution to the SHS signal is mostly averaged out due to symmetry. This is also borne out by the measured polarization (PPP > SSS), which is consistent with a closed surface and not with an open-ended structure. Furthermore, the large irregular shape of the dendritic voids makes it extremely unlikely to find a similar structure in close proximity on the femtosecond timescale as confirmed by MD simulations.

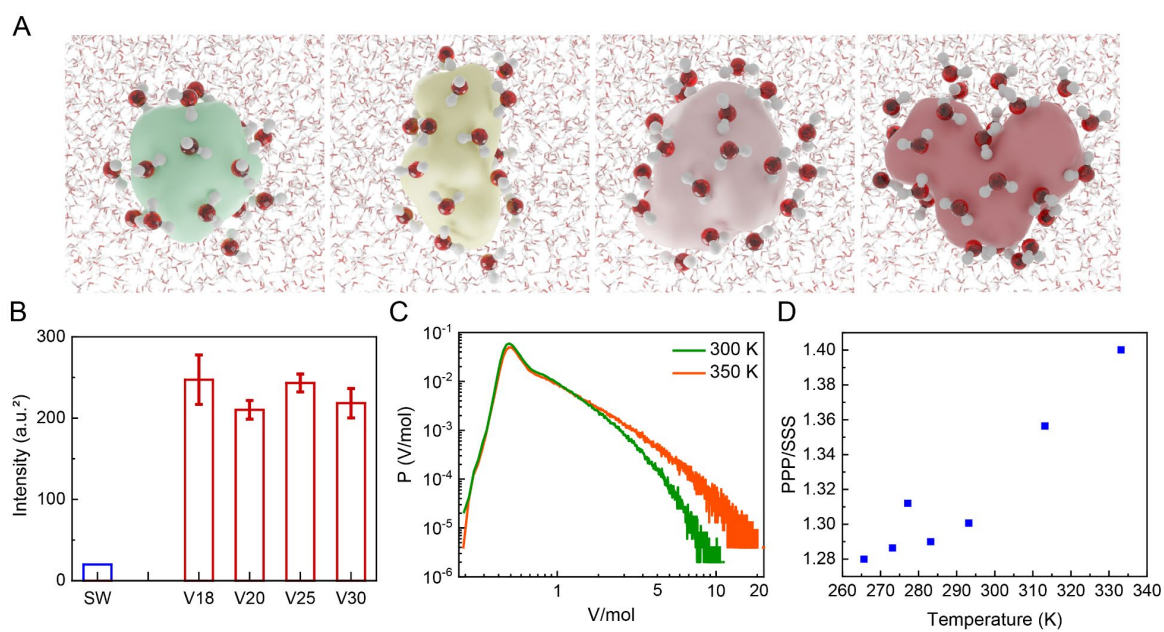

**Figure 3: SH intensity from transient nanoscale voids. A:** The voids from which the SH intensity was computed. **B:** Enhanced SH intensity around transient nanoscale voids. SH intensity per water molecule for water around voids (V18, V20, V25 and V30). The number refers to the number of water molecules surrounding the void. For comparison the intensity computed from single water (SW) is also shown. **C:** Probability distribution of void volumes normalized by the volume of a single water molecule (30 Å$^3$) for water at 300 and 350 K. An increase in the percentage of large voids is seen with an increase in temperature. **D:** The ratio of the measured coherent to incoherent SH intensity in the direction where **q** → 0° measured as a function of temperature. We observe an increase in the coherent response of 9 %.

Further evidence for our hypothesis is obtained by examining the temperature dependence of the SH intensity, which is done both computationally and experimentally. For uncorrelated voids the SH intensity depends linearly on the number of voids. If the number of voids increases with temperature, so does the SH intensity. Figure 3C displays the normalized void-volumes computed from the MD simulations at two different temperatures, 300 K (green curve) and 350 K (red curve). At 300 K the total volume of voids is determined by integrating this curve and amounts to ~ 3 vol% (equivalent to 1.5 M solute in water), and 20 % of these consisted of voids > 1 nm (equivalent to a concentration of 300 mM of dendritic voids in water). That this generates enough SH intensity to be detectable is understood from a comparison to previous experiments: Dilute electrolyte solutions with concentrations of 10 µM of simple ions generate sufficient detectable coherent fs-ESHS response, even though the computed derivation of the molecular ion induced tilt angle from the average tilt angle in liquid water was extremely small: 0.0076° at a distance of 1 nm from the ion [30]. This is a testimonial of the exquisite sensitivity of fs-ESHS to molecular anisotropy in the range of 1 - 10 nm. This type of sensitivity derives from the selection rule for structural anisotropy that all even-order nonlinear optical techniques have [44], and the absence of a background response since the emission occurs at a different wavelength than the probing beams. As the concentration of dendritic voids is much higher than 10 µM it is therefore certainly sufficient to create the coherent SH response, that is here measured.

Increasing the temperature (red curve, Fig. 3C) enhances the fraction of dendritic shaped voids by ~10%. Based on these predictions, we thus expect a comparable increase in the coherent SH intensity with increasing temperature. Figure 3D shows the experimental coherent contribution to the fs-ESHS response as quantified by the PPP/SSS intensity ratio measured for different temperatures. The coherent contribution increased by 9 % with temperature, in excellent agreement with the MD void predictions. Summarizing, the presence of voids as computed by MD simulations and the consequential enhanced polarization fluctuations lead to nanoscale heterogeneities in liquid water that were detected with fs-ESHS. Although fs-ESHS is a scattering technique and can therefore not directly be used to image a distribution of shapes, the polarization combinations, the directionality of the response, the intensity, and the temperature dependence are all consistent with the hypothesis that liquid water contains dendritic voids.

**Liquid water and its structure.** Bringing these results in connection with recent work, we discuss its relation to the uniform water picture [45], the 2-state water hypothesis [4,6] and the hypothesis of polymeric water [18]. Uniform liquid water at elevated temperatures has a decreased hydrogen bond strength [45], and consequently a lower density, viscosity, and cohesive energy [46,47]. Therefore water, with a decreased hydrogen bond strength has weaker orientational correlations and consequently, we expect a decrease in the coherent fs-ESHS response, which is inconsistent with the presented data.

Branched polymeric water as presented by Goddard [18], consists of clusters of water molecules that form three dimensional shapes reminiscent of branched polymers. If such structures indeed exist they will have to have an enclosing charge gradient along the normal direction, which seems unlikely but cannot be excluded.

Two state water has as a hall-mark the existence of more or less tetrahedrally oriented water. Such structures have been deduced by matching molecular dynamics simulations with temperature dependent X-ray scattering and spectroscopic data, with the spatially averaged structure factor as leading evidence [3,4,6]. Because the data analysis requires a manifold parameters and liquid water under the most useful temperature and pressure range to verify this model is inaccessible ('the no-mans' land of the phase diagram of water") [1], the exact details of the two state model is still a work in progress. As we remarked above, density fluctuations alone are not enough to generate a fs-ESHS response. Furthermore, if the coherent fs-ESHS response were due to a remainder of low and high density fluctuations at and above room temperature, the temperature dependence should have been reversed since the low and high-density structures have their most pronounced effect at low temperature in the super-cooled regime [16,17], and are virtually absent at elevated temperatures. This is the opposite of what we observe in Fig. 3 which shows an enhancement in both the number of voids as well as in the measured coherent SH intensity. Inversely, since the lifetime of the voids is on the timescale of hundreds of fs, they are unlikely to show up in the X-ray diffraction patterns in SAXS experiments [6]. Therefore, this study neither confirms nor disproves the two state water hypothesis.

**Conclusions**

In summary, fs-ESHS measurements of liquid $CCl_4$ and water in combination with MD simulation of neat water as well as a Lennard-Jones liquid can be understood to arise from transient ~ 300 fs cavities or voids in water, which lead to nanoscale structural anisotropies. The polarization fluctuations around the voids as revealed by the charge gradients and enhanced nanoscale polarizability, generates detectable coherent fs-ESHS emission, which increases with temperature. The structural anisotropies reported by the fs-ESHS experiment are different from those seen by X-ray based measurements, as the observables in both experiments are distinct and probe different length and time scales. However, both sets of experiments do confirm the presence of heterogeneities in liquid water, albeit in different manifestations. The fs-ESHS data shows that on femtosecond timescales transient nanoscale structures exist in liquid water at room temperature and above. Transient structural inhomogeneities in water as reported here will play important roles in chemistry and physics since they ensure non-uniform potential energy landscapes for any type of transition [48] or chemical reaction that might occur.

## Acknowledgments

This work was supported by the Julia Jacobi Foundation, and the European Research Council grant 616305.

## Supporting Information

Materials and Methods (S1), Coherent and incoherent contribution to SHS (S2), Exclusion of the possibility of an artifact (S3), Forward and backward scattering comparison (S4), MD simulations (D5), Voronoi voids and validation (S6), Linear scaling density functional theory (LS-DFT) calculations (S7), Nanoscale polarization calculations (S8).

## Author contributions

T.S., and N.D., performed experiments, N.D., T.S., and S.R. interpreted the experimental data. S.R., A.H., T.S., and N.D, wrote the manuscript. N.A., and E.P. performed the MD simulations, D.M.W. analyzed quantum-mechanical calculations, S.R. and A.H. conceived and supervised the work.

## Competing interests

The authors declare no competing interests.